\documentclass[prl,aps,assume,superscriptaddress,reprint]{revtex4-2}
\usepackage{graphicx}
\usepackage{dcolumn}
\usepackage{bm}
\usepackage{amsmath}
\usepackage{epstopdf}
\usepackage{natbib}
\usepackage{color}
\usepackage{lineno,hyperref}
\usepackage[normalem]{ulem}

\newcommand{\Rmnum}[1]{\expandafter\@slowromancap\romannumeral #1@}
\makeatother
\begin{document}
\title{\bf Large anomalous  Hall effect and \textit{A}-phase in hexagonal polar magnet Gd$_3$Ni$_8$Sn$_4$}

\author{ Arnab Bhattacharya }
\email{arnab.bh@saha.ac.in}
\affiliation{Condensed Matter Physics Division, Saha Institute of Nuclear Physics, A CI of Homi Bhabha National Institute, 1/AF, Bidhannagar, Kolkata 700064, India}
\author{Afsar Ahmed}
\affiliation{Condensed Matter Physics Division, Saha Institute of Nuclear Physics, A CI of Homi Bhabha National Institute, 1/AF, Bidhannagar, Kolkata 700064, India}
\author{Apurba Dutta}
\affiliation{Department of Physics, Indian Institute of Technology, Kanpur 208016, India}
\author{Ajay Kumar}
\affiliation{Ames National Laboratory, Iowa State University, Ames, Iowa 50011, USA}
\author{Anis Biswas}
\email{anis@ameslab.gov}
\affiliation{Ames National Laboratory, Iowa State University, Ames, Iowa 50011, USA}
\author{Yaroslav Mudryk}
\affiliation{Ames National Laboratory, Iowa State University, Ames, Iowa 50011, USA}
\author{ I. Das}
\affiliation{Condensed Matter Physics Division, Saha Institute of Nuclear Physics, A CI of Homi Bhabha National Institute, 1/AF, Bidhannagar, Kolkata 700064, India}
\begin{abstract}

While recent theoretical studies have positioned noncollinear polar magnets with $C_{nv}$ symmetry as compelling candidates for realizing topological magnetic phases and substantial intrinsic anomalous Hall conductivity, experimental realizations of the same in strongly correlated systems remain rare. Here, we present a large intrinsic anomalous Hall effect and extended topological magnetic ordering in Gd$_3$Ni$_8$Sn$_4$ with hexagonal $C_{6v}$ symmetry. Observation of topological Hall response, corroborated by metamagnetic anomalies in isothermal magnetization, peak/hump features in field-evolution of ac susceptibility and longitudinal resistivity, attests to the stabilization of skyrmion $A$-phase. The anomalous Hall effect is quantitatively accounted for by the intrinsic Berry curvature-mediated mechanism. Our results underscore polar magnets as a promising platform to investigate a plethora of emergent electrodynamic responses rooted in the interplay between magnetism and topology.




\end{abstract}

\maketitle

Integrating topology into magnetism has driven a fervent quest for novel quantum materials with exotic electronic and magnetic phases, aimed at unveiling \textcolor{black}{their} electrodynamic responses\cite{p43,p44,p18,p64}. \textcolor{black}{In particular}, topological magnets with non-trivial electronic band crossings proximate to Fermi level\cite{p47,p48,p51}, exhibit remarkable properties such as large magnetoresistance, enhanced intrinsic anomalous Hall conductivity (AHC) and anomalous Hall angle\cite{p51,p56,p54,p49,p50,p28,p29,p30}, thus positioning these materials as forerunners to explore the intertwining of topology and strong correlations\cite{p59}. \textcolor{black}{In} the real-space scenario, quantized topological defects in magnetic spin-lattice, such as skyrmions\cite{p24} and antiskyrmions\cite{p57}, characterized by finite scaler spin chirality $\chi_{ijk} = {\bm S_i} \cdot ({\bm S_j} \times {\bm S_k})$, where ${\bm S_n}$ are the neighbouring spins, induce an additional measurable component to the transverse resistivity ($\rho_{xy}$), $\rho_{xy}^T$, referred to as the topological Hall effect (THE)\cite{p25,p26,p58,p46}. This anomaly occurs as conduction electrons pick up a quantum mechanical Berry phase on coupling with the topological spin structure generated emergent gauge field, $B_{eff}$, in correlated systems\cite{p24}. However, the account of AHC, rooted in non-trivial electronic topology, has primarily centred on centrosymmetric collinear ferromagnets\cite{p48,p51}, limiting the exploration of coexisting topologically non-trivial incommensurate magnetic ordering and electronic bands. At this juncture, concurrently realizing these magnetic and electronic phenomena is crucial for advancing the understanding of hitherto unexplored interplay, thereby broadening the array of associated topological functionalities\cite{p60,p61,p54}.

In light of theoretical predictions by Bogdanov $et$ $al.$,\cite{p62} and Chang $et$ $al.$\cite{p63}, polar magnets with $C_{nv}$ crystal symmetry have emerged as a compelling platform for realizing this synergy. Contrasting B20 compounds, $C_{nv}$ symmetric materials with quenched orbital moment have a distinct advantage: the asymmetric Dzyaloshinskii-Moriya interaction restricts the magnetic modulation vector perpendicular to the polar axis, thereby suppressing the formation of competing conical phase\cite{p66}. This geometric confinement, combined with moderate easy-plane anisotropy, facilitates stabilising an extended Néel-type skyrmion phase, under an appropriate magnetic field along the polar axis\cite{p65,p67,p68}. Nevertheless, exploring skyrmion $A$-phase in bulk polar magnets has predominantly been confined to insulating lacunar spinels \cite{p67,p68}, limiting the investigation of associated electrodynamic responses\cite{p79}. This constraint underscores the need to broaden the material basis to include correlated bulk polar magnets with $A$-phase.

Here we address this pursuits by achieving extended $A$-phase down to lowest measured temperature along with substantial intrinsic AHC in Gd$_3$Ni$_8$Sn$_4$, a member of $R_3T_8$Sn$_4$ family \textcolor{black}{($R$ and $T$ being rare-earth and $3d$ elements, respectively)} of $C_{6v}$ point group, satisfying the prerequisites to host non-trivial magnetic ordering\cite{p62}, through detailed experimental study. Observed topological Hall response, corroborated by distinct metamagnetic step-like anomalies in isothermal magnetization and peak/hump features in the field evolution of ac-susceptibility, is suggestive of the field-induced stabilization of $A$-phase. Specific heat measurements confirm a long-range spin-modulated ground state. Furthermore, the Berry curvature mediated large intrinsic AHC underscores the $R_3T_8$Sn$_4$ family of metallic polar magnets as a unique backdrop for simultaneously realising electronic and magnetic topological functionalities.

\begin{figure}[t]
	\begin{center}
		\includegraphics[width=0.49\textwidth]{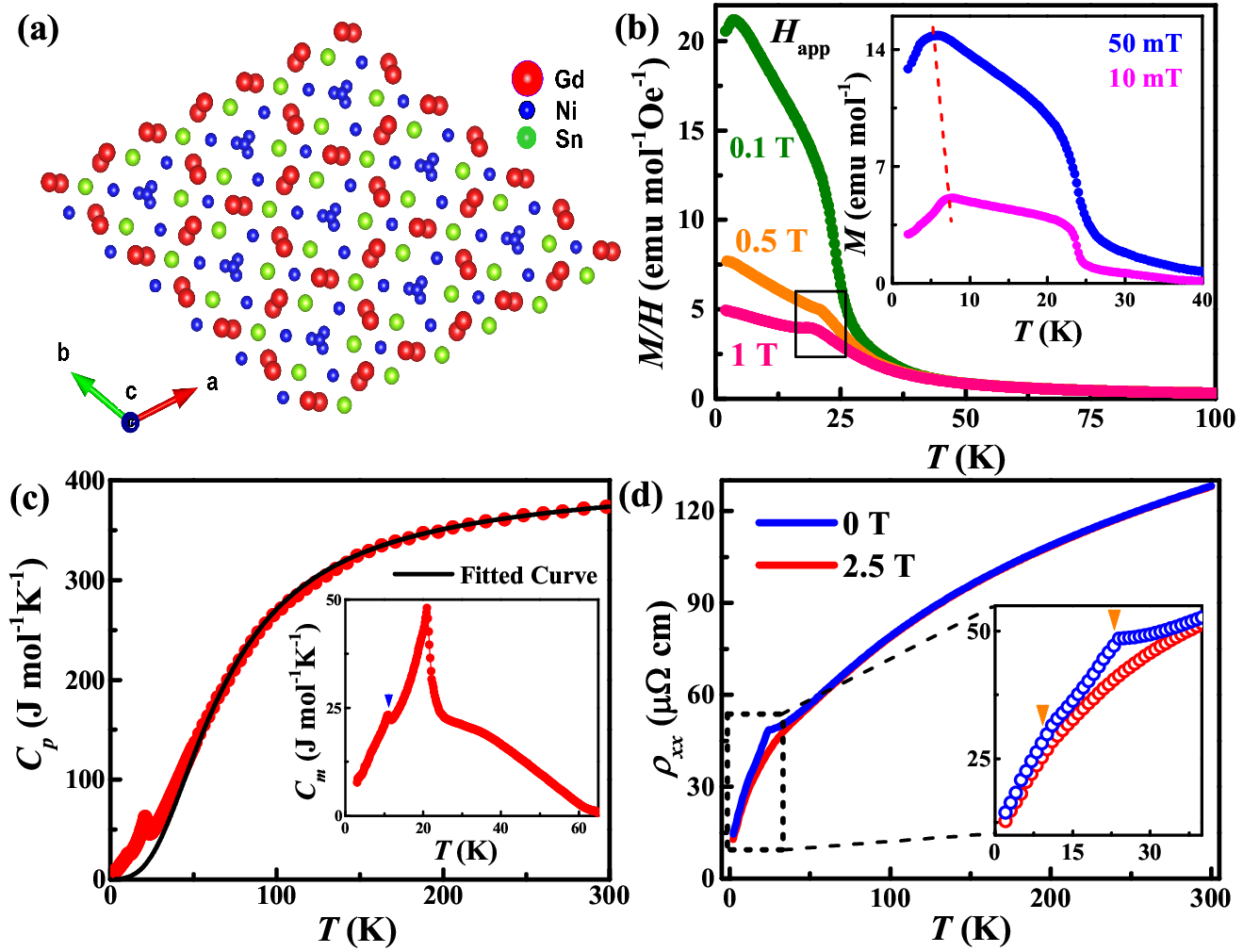}
	\end{center}
	\caption{(a) Crystal structure (top view) of Gd$_3$Ni$_8$Sn$_4$. (b) Magnetic susceptibility $\chi(T)$ curves under applied fields of 0.1, 0.5 and 1 T. The magnetization kink is highlighted in black box and the inset shows low-field $M(T)$ curves. (c) Zero-field heat capacity $C_p$. Black line is the non-magnetic lattice model comprising of electronic contribution and Debye function. Inset shows the 4$f$ magnetic contribution to $C_p$. (d) Temperature variation of $\rho_{xx}$ under $H_{app}$ of 0 and 2.5 T. Inset shows the distinct kinks as mentioned in the text. }
	\label{F1}
\end{figure}

Rietveld refinement of room temperature powder X-ray diffraction pattern of the \textcolor{black}{samples prepared by conventional arc-melting technique} confirms the single-phase nature of Gd$_3$Ni$_8$Sn$_4$, which crystallizes in hexagonal polar structure (space group $P6_3mc$) (See SF1 and Table.ST1 of supplementary material \cite{R30}). The lattice parameters are in good agreement with the previous report\cite{p1}. The crystal structure is a distorted derivative of the BaLi$_4$ structure (space group $P6_3/mmc$), resulting in the splitting of atomic positions and broken inversion ($\mathcal{I}$) symmetry, with the polar axis along c-axis (Fig.\ref{F1}(a))\cite{p2,p9}.


To probe the magnetic properties of Gd$_3$Ni$_8$Sn$_4$, we performed \textcolor{black}{dc thermomagnetic} measurements. Figure.\ref{F1}(b) illustrates the temperature variation of \textcolor{black}{dc magnetic} susceptibility ($\chi_{dc}= M/H$) under various applied fields ($H_{app}$), revealing a transition from paramagnetic to magnetically-ordered state at $T_C$ = 24 K under $H_{app}$ = 10 mT (inset of Fig.\ref{F1}(b)). Gd$_3$Ni$_8$Sn$_4$ follows a Curie-Weiss behaviour with effective paramagnetic moment ($\mu_{eff}$) of 8.11$\mu_B$/Gd$^{+3}$ ions, slightly larger than theoretical value of 7.94$\mu_B$ for free Gd$^{+3}$ ions\cite{p5} and a positive Weiss temperature $\Theta_{CW}\approx$ 18 K, reflecting the dominant ferromagnetic (FM) correlations. This small difference of $\mu_{eff}$ from \textcolor{black}{theoretical} value might arise from the involvement of itinerant conduction electrons. A notable decrease in magnetization arises at ($T_K \sim$) 7 K for small $H_{app}$ of 10 mT, persisting up to 0.1 T, with $T_K$ shifting to lower temperatures as $H_{app}$ increases (inset Fig.\ref{F1}(b)). However, the overlapped zero-field cooled (ZFC) and field-cooled (FC) thermomagnetic curves down to 2 K, along with the absence of an upward peak-shift in the real component of ac-susceptibility $\chi'(T)$ across increasing frequencies \textcolor{black}{(Fig.SF2(b) of \cite{R30})}, collectively rules out the presence of a magnetically frustrated state \cite{p1}. This suggests the development of an antiferromagnetic (AFM) component below $T_K$. However, for this instance, the small $H_{app}$ of 0.15 T is strong enough to melt away this AFM component. Along with the low-temperature anomaly, adjacent to $T_C$, a distinct cusp emerges in the intermediate $H_{app}$ of 0.5 and 1 T, followed by an increase in magnetization as temperature decreases (Fig.\ref{F1}(b) and \textcolor{black}{Fig.SF2(a) of \cite{R30}}). This cusp broadens with increasing $H_{app}$, indicating a complex magnetic ordering rather than purely FM state.\cite{p3,p4}.

Figure.\ref{F1}(c) illustrates specific heat $C_p$ as a function of temperature in the range of 2-300 K. The saturated $C_p$ at room temperature $\sim 373.2$ Jmol$^{-1}$K$^{-1}$ is in good alignment with the Dulong-Petit limit of $C_p = 3nR$ = 374.13 Jmol$^{-1}$K$^{-1}$, where $R$ is the ideal gas constant and $n= 15$ is the number of atoms per formula unit. Above 60 K, the $C_p$ can be well-accounted by the band electron component ($C_{el}$) and phononic Debye model ($C_{Deb}$) as, $C_{mod} = C_{el} + C_{Deb} = \gamma_{el}\cdot T + 9N_DR(\frac{T}{\theta_D})^3\int_{0}^{\theta_D/T}\frac{x^4e^x}{(e^x-1)^2}dx$ \cite{p10,p69}. Here, $\gamma_{el}$, $N_D$ and $\theta_{D}$ are Sommerfield coefficient, number of Debye oscillator and Debye temperature, respectively. The model fitting of $C_p(T)$, as in Fig.\ref{F1}(c), estimates $\theta_{D} =$ 269.3 K and $\gamma_{el} =$ 56 mJmol$^{-1}$K$^{-2}$. To discern the $4f$ magnetic contribution ($C_{m}$) to $C_p$, we subtracted the extrapolated model curve to low temperatures, $C_{m} = C_{p} - C_{mod}$ (Inset Fig.\ref{F1}(c)). The pronounced $\lambda$-peak at $T \sim$ 21 K, \textcolor{black}{slightly below} $T_C$, attests to the long-range nature of the magnetic ordering while a subsequent anomaly at lower temperatures corroborates with $T_K$. Accounting in the mean-field theory concerning the behavior of $C_{m}$ for localized moment systems\cite{p11,p12}, equal moment arrangement of Gd$^{+3}$ ions with $J$ = 7/2 should yield $C_{m}^{\mathrm{MFT}}$ = $\frac{5S(S+1)}{2J^2 + 2J + 1}R$ = 20 Jmol$^{-1}$K$^{-1}$. Contrary to this prediction, the observed $C_{m}$ peak ($\sim 16$ Jmol$^{-1}$K$^{-1}$/Gd$^{+3}$) is much smaller than $C_{m}^{\mathrm{MFT}}$, \textcolor{black}{indicating} a more complex modulated magnetic ordering than conventional equal-moment magnetic ground state\cite{p13,p14,p15,p16,p17}, endorsing the magnetization results.

Figure.\ref{F1}(d) depicts the temperature dependence of longitudinal resistivity ($\rho_{xx}$) under 0 and 2.5 T $H_{app}$ with a residual resistivity ratio (=$\rho_{300 \mathrm{K}}/\rho_{2 \mathrm{K}}$) of 8.8 indicating the high quality of prepared samples. Notably, $\rho_{xx}(T)$ under 0 T exhibits a distinct peak at $T_C$, marking the onset of long-range ordering, followed by a rapid decrease due to the suppression of spin-disorder scattering. Above $T_C$, $\rho_{xx}(T)$ shows an upturn, \textcolor{black}{attributable to enhanced spin-fluctuations due to short-range Gd-4$f$ correlations prior to long-range ordering}\cite{p15,p16,p17,p20,p18}, while it also captures the second transition at $T_K$. However, the peak at $T_C$ melts away with the application of 2.5 T, establishing an intricate correlation between magnetic ordering and transport properties.

\begin{figure}[h]
	\begin{center}
		\includegraphics[width=0.49\textwidth]{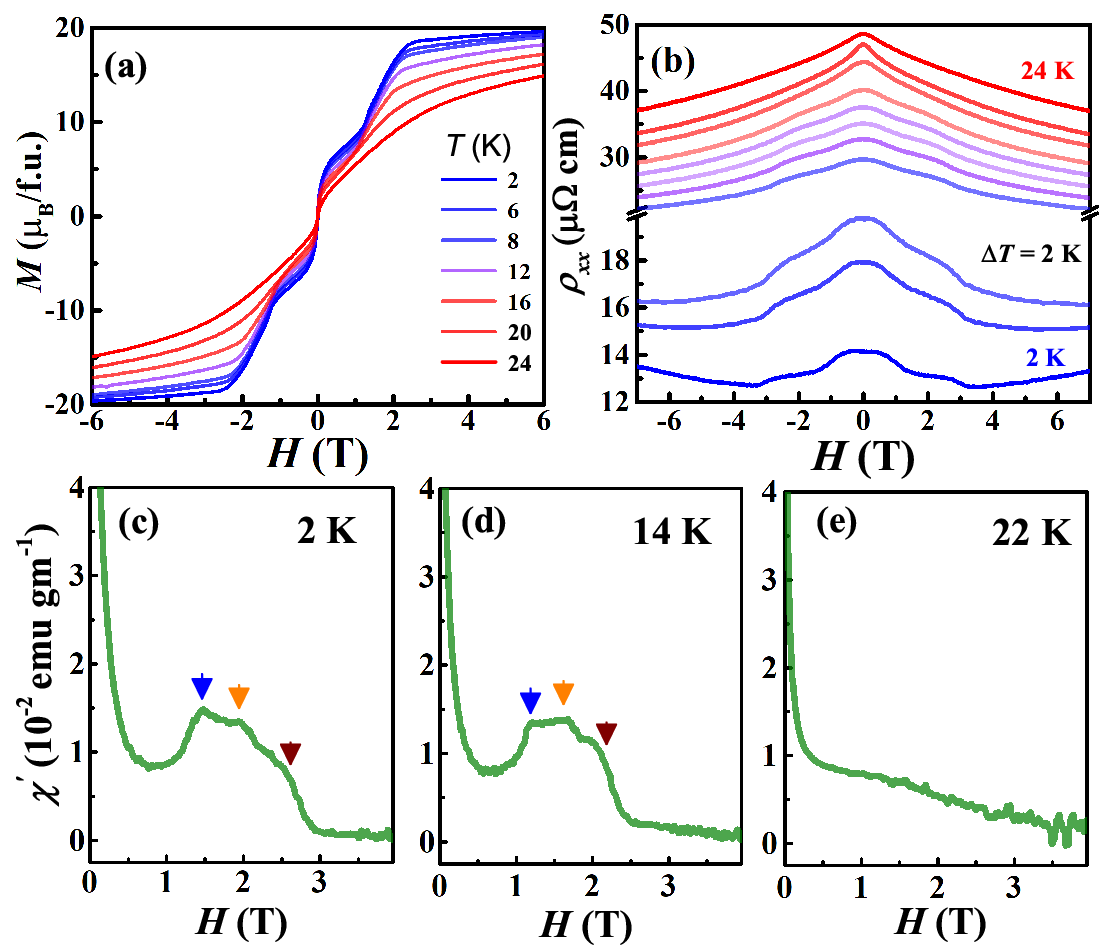}
	\end{center}
	\caption{(a)  Isothermal magnetization $M$ as a function of applied magnetic field. (b) Isothermal $\rho_{xx}(H)$ showcasing anomaly associated with phase transition below $T_C$. (c)-(e) Field dependant ac-susceptibility measured at different temperatures.}
	\label{F2}
\end{figure}

To elucidate the field-induced modifications in magnetic ordering, we investigate the isothermal dc-magnetization $M(H)$  (in Fig.\ref{F2}(a)). The magnetization profile reveals distinct metamagnetic step-like anomalies before attaining the field-polarized state. Notably, these metamagnetic features weaken as $T$ increases and approaches $T_C$ but persist on either side of $T_K$, implying a robust field-induced stabilization of complex magnetic ordering\cite{p18,p21,p22}. Intriguingly, isothermal $\rho_{xx}(H)$ as illustrated in Fig.\ref{F2}(b), displays a pronounced change of slope at fields corresponding to the metamagnetic transitions, driven by significant alterations of carrier lifetime due to domain wall scattering associated with these magnetic transitions\cite{p18,p21,p22}. To gain further insight into these metamagnetic transitions we analyzed the field-evolution of ac-susceptibility, $\chi'(H)$, which has been extensively employed to characterize various topologically trivial and non-trivial magnetic phases due to associated changes in the energy landscape\cite{p35,p34,p33,p32}. Figure.\ref{F2}(c)-(e) illustrates the $\chi'(H)$ under an applied ac frequency of 333 Hz at different temperatures. Pronounced peak/dip anomalies (marked by arrows) are evident at fields corresponding to the metamagnetic transitions in isothermal dc-magnetization and $\rho_{xx}(H)$ at 2 K and 14 K but vanish completely at 22 K. Following these anomalies, $\chi'(H)$ exhibit monotonous behavior as Gd$_3$Ni$_8$Sn$_4$ attains a field-polarized state. It is noteworthy, similar features in $\chi'(H)$ have been reported for Gd$_2$PdSi$_3$\cite{p23} and $D_{2d}$ Heusler alloys\cite{p42,p77}, across their skyrmion and anti-skyrmion phase pocket, respectively. The persistence of these peak/dip anomalies across different temperatures in $\chi'(H)$ underscores the field-driven stabilization of a robust extended topological $A$-phase \textcolor{black}{which subsequently melts to a field-polarized state under high $H_{app}$.}

\begin{figure}[t]
	\begin{center}
		\includegraphics[width=0.48\textwidth]{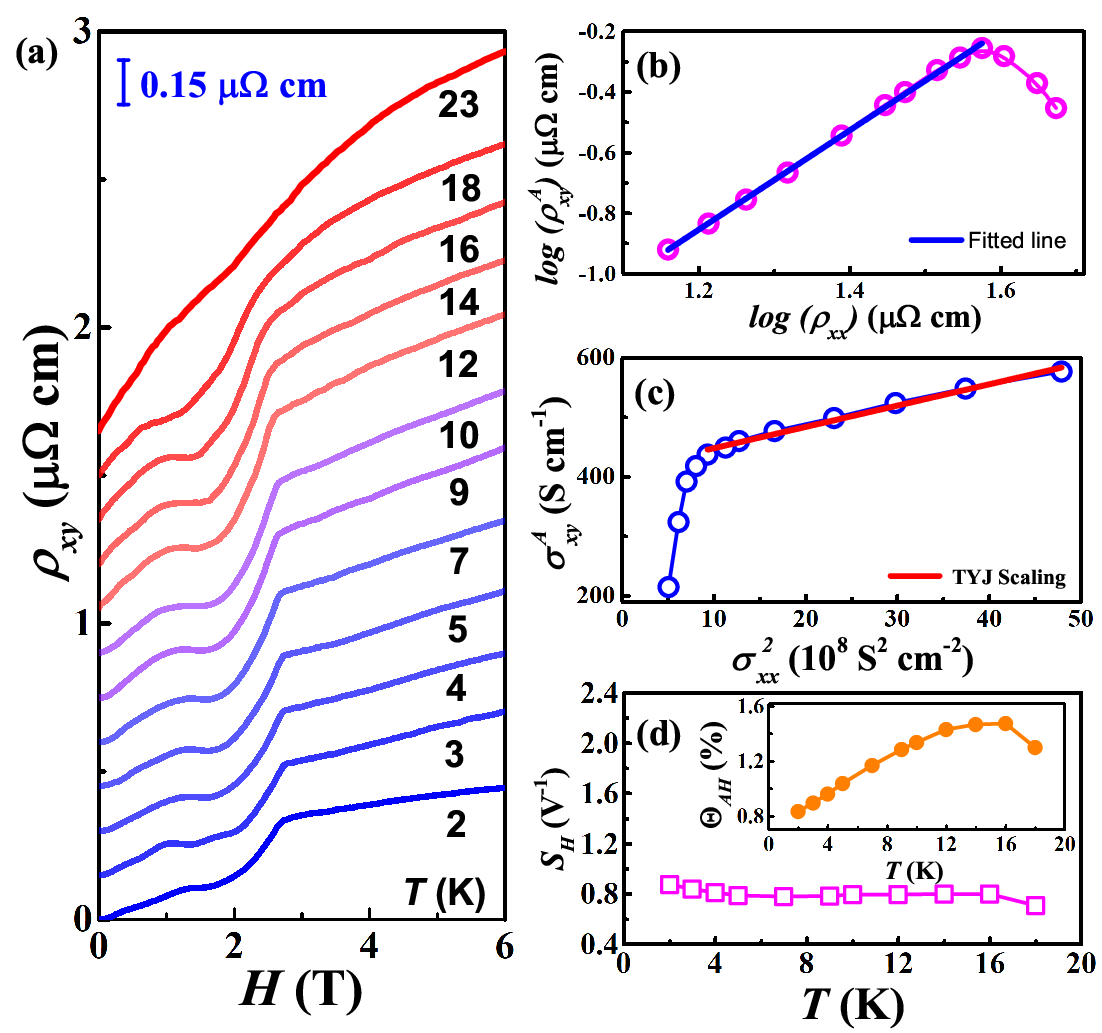}
	\end{center}
	\caption{(a) Field dependant isothermal $\rho_{xy}$ at various $T$. For better visualization, an offset has been created along $\rho_{xy}$ axis. (b) log($\rho_{xy}^A$) versus log($\rho_{xx}$) with the slope of $\alpha \approx 1.63$  (c) $TYJ$ scaling plot for AHC. (d) $T$ variation of the anomalous Hall coefficient $S_H$. Inset depicts the $T$ dependency of anomalous Hall angle $\Theta_{AH}$}
	\label{F3}
\end{figure}

\begin{figure}[t]
	\begin{center}
		\includegraphics[width=0.48\textwidth]{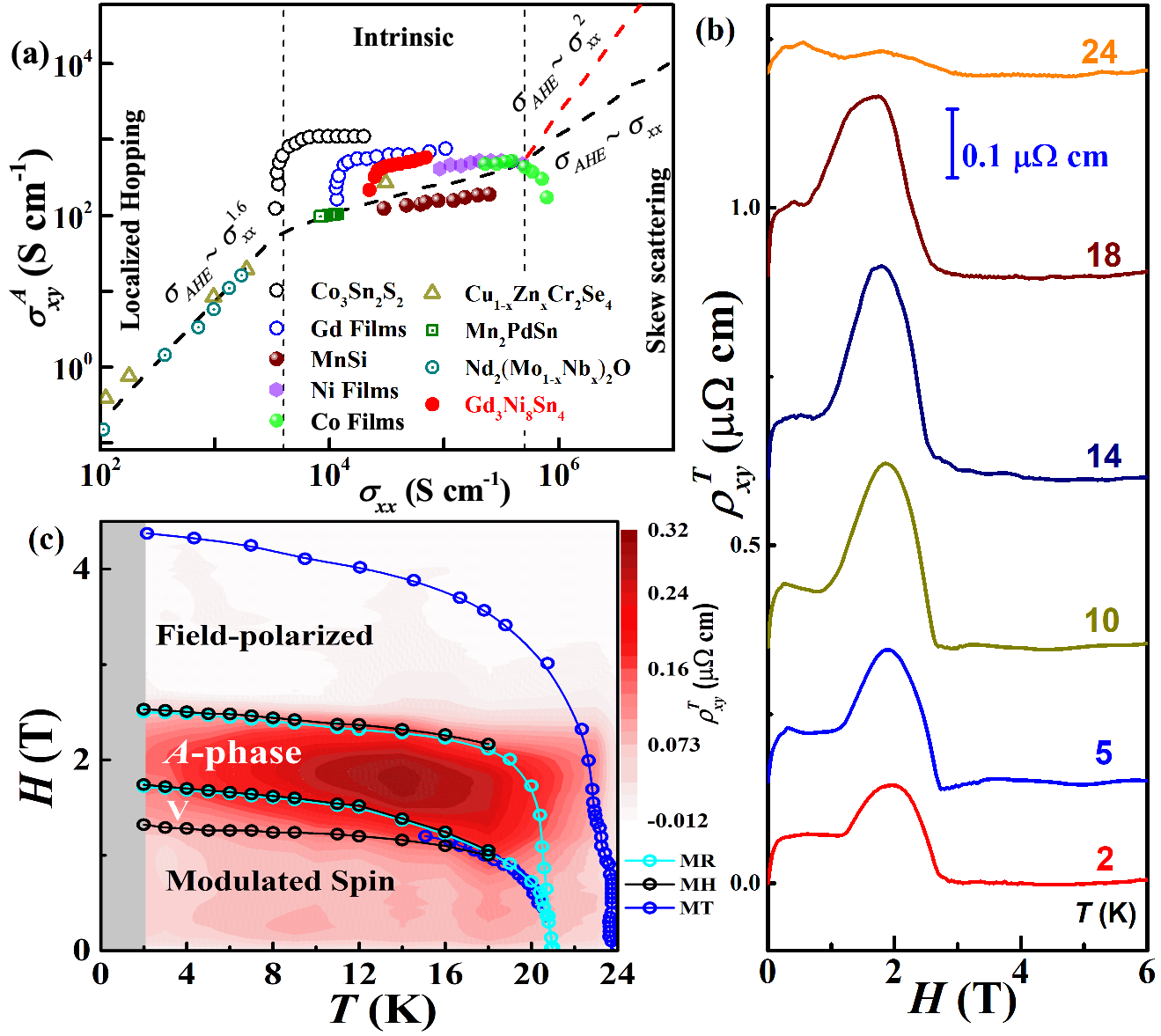}
	\end{center}
	\caption{(a) Universal plot of $\sigma_{xy}^A$ with $\sigma_{xx}$, illustrating the intrinsic regime for Gd$_3$Ni$_8$Sn$_4$ along with other magnetic conductors (b) Field dependency of $\rho_{xy}^T$, suggesting the formation of topological magnetic ordering. A vertical offset has been created for clear presentations. (c) Contour plot of $\rho_{xy}^T$ in $H$-$T$ diagram with the phase boundaries obtained from the derivative of $\chi_{dc}(T)$, isothermal $M(H)$ and $\rho_{xx}(H)$. }
	\label{F4}
\end{figure}

Building on the established correlation between topological $A$-phase and $\rho_{xy}(H)$\cite{p24,p25,p26}, we turn to the Hall transport data. Figure.\ref{F3}(a) illustrates the field dependence of $\rho_{xy}$ at various $T$. For magnetic conductors with non-trivial magnetic orderings, $\rho_{xy}$ is empirically expressed as, $\rho_{xy} = \rho_{xy}^N + \rho_{xy}^A + \rho_{xy}^T$, where $\rho_{xy}^N = R_0H$ and $\rho_{xy}^A = R_sM$ are normal and anomalous Hall resistivity with coefficients $R_0$ and $R_S$, respectively. The positive slope of $\rho_{xy}(H)$ attests to holes as majority charge carriers with a carrier density of $n_0 \sim 2 \times 10^{22}$ cm$^{-3}$ at $T$ = 2 K, inferred from the relation $n_0 = -1/|e|R_0$. $\rho_{xy}^A$ is obtained by high field extrapolation of $\rho_{xy}(H)$, where magnetization attains saturation, to zero field (Fig. SF3(a) of \cite{R30}). \textcolor{black}{Figure.SF3(b) of \cite{R30}} illustrates a monotonic temperature variation of $\rho_{xy}^A$. In a general framework, AHE can stem from reciprocal space originated Berry curvature driven intrinsic or asymmetric scattering induced extrinsic side-jump (\textit{sj})/skew-scattering (\textit{sk}) mechanisms or a combination of both, where depending upon origin, $\rho_{xy}^A$ scales differently with $\rho_{xx}$ \cite{p27}. In Fig.\ref{F3}(b), the log($\rho_{xy}^A$) vs. log($\rho_{xx}$) curve is fitted using the linear relation $\rho_{xy}\propto \rho_{xx}^\alpha$ which yields $\alpha \approx 1.63$, validating the dominant contribution from intrinsic or $sj$ mechanism in AHE\cite{p31}. For a better insight into the dominating contribution, we derive the AHC, $\sigma_{xy}^A$, from total Hall conductivity $\sigma_{xy} \approx \rho_{xy}/\rho_{xx}^2$, employing method analogous to obtaining $\rho_{xy}^A$. The near temperature-invariant nature exhibited by $\sigma_{xy}^A$ (Fig.SF3(b) of \cite{R30}), with a value of $\sim 577$ S cm$^{-1}$ at $T$ = 2 K, is suggestive of the intrinsic origin of AHE\cite{p30,p29,p28}. To quantify the different contributions, we adopt the Tian-Ye-Jin scaling relation for $\sigma_{xy}^A$ $i.e.$, $\sigma_{xy}^A = -\kappa\sigma_{xx0}^{-1}\sigma_{xx}^2 - b = -a\sigma_{xx}^2 - b$ \cite{p31}. Here, $\sigma_{xx0}$ and $b = \rho_{xy}^A/\rho_{xx}^2$ corresponds to residual longitudinal conductivity and intrinsic AHC $\sigma_{xy,int}^A$, respectively. The scaling expects a linear relation of $\sigma_{xy}^A$ with $\sigma_{xx}^2$, illustrated in Fig.\ref{F3}(c), with the intercept yielding $b \approx$ 413 S cm$^{-1}$. The deviation from scaling as $T$ approaches $T_C$ can originate from the broadening of the Fermi-Dirac distribution\cite{p54,p48}. Thus, $\sigma_{xy,int}^A$ accounts to 71$\%$ of $\sigma_{xy}^A$ at $T$ = 2 K, underpinning the dominance of Berry curvature-driven contribution. However, at low temperatures $sj$ conductivity, $\sigma_{xy,sj}^A$, remains closely intertwined with $\sigma_{xy,int}^A$ due to reduced phonon scattering, complicating individual quantification in the absence of theoretical framework. While the estimation of the order of magnitude of $\sigma_{xy,sj}^A$ shows that it is two orders smaller than $\sigma_{xy,int}^A$\cite{p27,p36}, confirming the dominance of intrinsic mechanism in AHE.


Figure.\ref{F3}(d) illustrates temperature variation of the anomalous Hall coefficient $S_H$ ($=\sigma_{xy}^A/M$), quantifying the sensitivity of anomalous Hall current to magnetization. Notably, $S_H$ remains invariant with temperature for $\sim$ 0.8 V$^{-1}$, confirming the robustness and insensitivity of AHE to impurity scattering\cite{p28}. On employing the second characteristics parameter, $i.e.$, the anomalous Hall angle, $\Theta_{AH} = \sigma_{xy}/\sigma_{xx}(\%)$, a monotonic increase with temperature is observed (inset of Fig.\ref{F3}(d)). Figure.\ref{F4}(a) shows the variation of $\sigma_{xy}^A$ with $\sigma_{xx}$ of Gd$_3$Ni$_8$Sn$_4$ along with other magnetic conductors. Here it is well within the intrinsic limit and remains invariant with $\sigma_{xx}$, anchoring the Berry curvature rooted picture of AHE\cite{p37,p38}.

Leveraging the intrinsic origin of AHE, $R_S$ is formulated as $R_S = \gamma \rho_{xx}^2 $. Hence, to obtain $R_s$, we modeled the transverse resistivity as $(\rho_{xy}/H) = R_0 + \gamma (\rho_{xx}^2 M)/H$. The $(\rho_{xy}/H)$ versus $(\rho_{xx}^2 M)/H$ plot showcases a good linear relation in the field-polarized state \textcolor{black}{(Fig.SF3(c) of \cite{R30})}, with the slope and intercept yielding $\gamma$ and $R_0$, respectively. Figure.\ref{F4}(b) shows the field dependency of $\rho_{xy}^T$ at various $T$, obtained by subtracting the calculated ($\rho_{xy}^N + \rho_{xy}^A$) from $\rho_{xy}$ \textcolor{black}{(Fig.SF3(a),(d) of \cite{R30})}. $\rho_{xy}^T$ attains a maximum value of 0.32 $\mu\Omega$cm at 14 K ($\sigma_{xy}^T \sim 253$ Scm$^{-1}$) and gradually diminishes with increasing temperature. The evolution of $\rho_{xy}^T$, as $T$ approaches $T_C$, attests to the consistency with the scaler spin-chirality model, suggesting a significant contribution of the molecular field from 4$f$-moment on conduction electrons through $f$-$d$ coupling\cite{p41}. Figure.\ref{F4}(c) shows that the peak positions in the derivative of isothermal magnetization $M(H)$, longitudinal resistivity $\rho_{xx}(H)$ and temperature-dependant magnetization $M(T)$ distinctly mark the boundaries of Field-polarized, $A$-phase and modulated spin phase along with a metastable V-phase (see Fig.SF4 of \cite{R30}). Accounting in the theoretical model proposed by Rowland \textit{et al,}\cite{p82} and broken-$\mathcal{I}$ along $z$-axis, we speculate, a dominant Rashba spin-orbit coupling stabilizes a sliver square skyrmion lattice (V-phase) adjacent to dominant $A$-phase. Overlaying $\rho_{xy}^T$ contour plot with the $H$-$T$ phase diagram shows that the enhanced $\rho_{xy}^T$ exclusively appears in the $A$-phase and V-phase region, suggesting $H$-driven evolution of the spin-modulated ground state to topological spin textures. It is \textcolor{black}{worth mentioning}, a shoulder-like anomaly in $\rho_{xy}^T$ adjacent to the peak in the $A$-phase is attributable to the complex modulated ground state magnetic ordering \cite{p70}.

For a comprehensive perspective, the maximum value of $\rho_{xy}^T$ ($\sim$0.32 $\mu\Omega$cm) is comparable with large wavelength spin-modulated systems like cubic FeGe (0.16 $\mu\Omega$cm)\cite{p39}, polar chiral magnet GdPt$_2$B (0.14 $\mu\Omega$cm)\cite{p3} and $D_{2d}$ Heusler alloy Mn$_{1.4}$Pt$_{0.9}$Pd$_{0.1}$Ga (0.4 $\mu\Omega$cm)\cite{p42}. This contrasts sharply with Gd-based centrosymmetric compounds like Gd$_2$PdSi$_3$\cite{p18}, Gd$_3$Ru$_4$Al$_{12}$\cite{p21} and GdRu$_2$Si$_2$\cite{p22} with short-period spin modulation wavelength squeezing the skyrmions to smaller sizes. Accounting in the theory of THE\cite{p71}, the real space Berry curvature primarily contributes to the limit $l<a_t$, where $l= \frac{\hbar}{n_0e^2}(3\pi^2n_0)^\frac{1}{3}\sigma_{xx0} $ and $a_t$ are the mean-free path of electrons and size of skyrmion, respectively. To verify this, we performed a semiquantitative analysis of THE, drawing on the established topological nature of the $A$-phase. In continuum approximation for smooth winding spin texture, the $B_{eff}$ is defined as $B_{eff} = h/ea_{t}$ and is related to $\rho_{xy}^T$ by spin polarization $P$ of conduction electron and normal Hall coefficient $R_0$, $\rho_{xy}^T = B_{eff}PR_0$ \cite{p24}. The conduction electron polarization can be crudely estimated as $P = M_{spo}/M_S$, where $M_{spo}$ is the ordered moment in the $A$-phase, by which we arrive at $P$ = 0.8. Employing the maximum observed $\rho_{xy}^T$ at 14 K, we derive $B_{eff} \sim$ 3.7 T corresponding to $a_t \sim$ 34 nm $>l \approx$ 10-15 nm at $\rho_{xx} \sim $  35 $\mu\Omega cm$, acknowledging the real-space $A$-phase driven picture of THE while being consistent with the consensus of stabilizing large-sized skyrmion (20-200 nm) by DM interaction\cite{p64,p81}.


In conclusion, this comprehensive experimental study establishes metallic polar magnets as a conducive backdrop for realizing the topological $A$-phase and exploring the reciprocal-space mediated emergent electrodynamic responses. The topological Hall effect, supplemented by peak/dip behaviour in field-dependent ac-susceptibility, is attributed to the stabilization of an extended topological magnetic quasiparticle phase of N\'eel-type skyrmion, as per theoretical predictions for systems exhibiting $C_{nv}$ symmetry, and a potential narrow square skyrmion phase. These results validate the theoretically-laid phase diagram of skyrmion-hosting polar magnets\cite{p82}. Additionally, the quadratic relation between $\rho_{xy}^A$ and $\rho_{xx}$ quantitatively attests to the role of reciprocal-space Berry curvature in generating the substantial AHC, resonating with the possibility of Weyl nodes in moderately $\mathcal{I}$-broken systems\cite{p63,p76}. These findings underscore the broken-$\mathcal{I}$ inherent to the $R_3T_8$Sn$_4$ family as a compelling framework to explore the non-trivial bulk electronic band topology and non-collinear magnetic ordering. Our findings advocate further experimental and theoretical investigation focused on correlated polar magnets to unravel the intricate interplay of reciprocal-space-mediated electromagnetic fields and magnetization dynamics to leverage topological functionalities.

\textit{Acknowledgement}\indent A. Bhattacharya and A. Ahmed acknowledge SINP, India and the Department of Atomic Energy (DAE), Government of India for their Fellowship. Work at the Ames National Laboratory was supported by the Division of Materials Science and Engineering of the Office of Basic Energy Sciences, Office of Science of the U.S. Department of Energy (D.O.E). Ames National Laboratory is operated for the U.S. DOE by Iowa State University of Science and Technology under Contract No. DE-AC02-07CH11358. We thank Prof. E. V. Sampathkumaran, TIFR, for his valuable suggestions.

\bibliography{ref}
\end{document}